\documentclass[onecolumn,12pt]{article}
\usepackage{graphicx}
\begin{document}
\title{Neutrino Mass Matrix from Seesaw Mechanism with Heavy Majorana Mass Matrix Subject to Texture Zero and Invariant Under a Cyclic Permutation}
\date{}
\maketitle
\author{\small Asan Damanik$^{(1,2)}$, Mirza Satriawan$^{(2)}$, Pramudita Anggraita$^{(3)}$, Arief Hermanto$^{(2)}$, and Muslim$^{(2)}$ }\\
\begin{center}
$^{(1)}$Department of Physics, Sanata Dharma University,\\ 
Kampus III USD Paingan, Maguwoharjo, Sleman, Yogyakarta, Indonesia\\
\end{center}
\begin{center}
$^{(2)}$Department of Physics, Gadjah Mada University,\\  
Bulaksumur, Yogyakarta, Indonesia.\\
\end{center}
\begin{center}
$^{(3)}$National Nuclear Energy Agency (BATAN), Jakarta, Indonesia.\\
\end{center}
\begin{center}
E-mail:d.asan@lycos.com, mirza@ugm.ac.id, pramudita@batan.go.id
\end{center}

\abstract{We evaluate the predictive power of the neutrino mass matrices arising from seesaw mechanism with heavy Majorana mass matrices subject to texture zero and satisfy a cyclic permutation invariant form to the solar neutrino mixing phenomena.  From eight possible patterns of heavy Majorana neutrino mass matrix, we found that there is no heavy Majorana neutrino mass matrix to be invariant in form under a cyclic permutation.  But, by imposing an additional assumption that at least one of the $2\times2$ sub-matrices of heavy Majorana neutrino mass matrix inverse having zero determinant, we found that only two of the eight possible patterns for heavy Majorana neutrino mass matrices to be invariant under a cyclic permutation.  One of the two invariant heavy Majorana neutrino mass matrices could produces neutrino mass matrix $M_{\nu}$ that can be used to explain the neutrino mixing phenomena for both solar and atmospheric neutrinos qualitatively.}

\section{Introduction}
Even though the Glashow-Weinberg-Salam (GWS) model for electroweak interaction (standard model for electroweak interaction) which is based on $SU(2)_{L}\otimes U(1)_{Y}$ gauge symmetry group (see for example Peskin and Schroeder \cite{Peskin}) has been in success phenomenologically, but it still far from a complete theory because the GWS model could not explain many fundamental problems such as neutrino mass problem, and fermions (lepton and quark) mass hierarchy \cite{Fukugita}. For more than two decades, solar neutrino flux measured on Earth has been much less than predicted by the solar model \cite{Pantaleone}. Solar neutrino deficit can be explained if neutrino undergoes oscillation during its propagation to earth.  Neutrino oscillation is the change of a one kind of neutrino flavor to another one during its propagation through vacuum or matter. Neutrino oscillation implies that neutrinos have a non-zero mass or at least one of the three known neutrino flavors has a non-zero mass and neutrino mixing does exist.
Recently, there is a convincing evidence that neutrinos have a non-zero mass.  This evidence was based on the experimental facts that both solar and atmospheric neutrinos undergo oscillation \cite{Fukuda}. These facts are in contrast with the GWS model that taking the neutrinos to be massless.

A global analysis of neutrino oscillations data gives the best fit value to solar neutrino mass-squared differences \cite{Gonzales-Garcia}
\begin{eqnarray}
\Delta m_{21}^{2}=(8.2_{-0.3}^{+0.3})\times 10^{-5}~eV^2~
 \label{eq:1}
\end{eqnarray}
with
\begin{eqnarray}
\tan^{2}\theta_{21}=0.39_{-0.04}^{+0.05},
 \label{eq:2}
\end{eqnarray}
and for the atmospheric neutrino mass-squared differences
\begin{eqnarray}
\Delta m_{32}^{2}=(2.2_{-0.4}^{+0.6})\times 10^{-3}~eV^2~
 \label{eq:3}
\end{eqnarray}
with
\begin{eqnarray}
\tan^{2}\theta_{32}=1.0_{-0.26}^{+0.35},
 \label{eq:4}
\end{eqnarray}
where $\Delta m_{ij}^2=m_{i}^2-m_{j}^2~ (i,j=1,2,3)$ with $m_{i}$ as the neutrino mass eigenstates basis $\nu_{i}~(i=1,2,3)$ and $\theta_{ij}$ is the mixing angle between $\nu_{i}$ and $\nu_{j}$.  The relation between neutrino mass eigenstates and neutrino weak (flavor) eigenstates basis $(\nu_{e},\nu_{\mu},\nu_{\tau})$ is given by
\begin{eqnarray}
\bordermatrix{& \cr
&\nu_{e}\cr
&\nu_{\mu}\cr
&\nu_{\tau}\cr}=V\bordermatrix{& \cr
&\nu_{1}\cr
&\nu_{2}\cr
&\nu_{3}\cr},
 \label{eq:5}
\end{eqnarray}
where $V$ is the mixing matrix.

To explain a non-zero neutrino mass-squared differences and neutrino mixing, several models for neutrino mass and its underlying family symmetries, and the possible mechanism for generating a neutrino mass have been proposed \cite{Leontaris}.  One of the interesting mechanism to generate neutrino mass is the seesaw mechanism \cite{Minkowski}, in which the right-handed neutrino $\nu_{R}$ has a large Majorana mass $M_{N}$ and the left-handed neutrino $\nu_{L}$ is given a mass through leakage of the order of $~(m^2/M_{N})$ with $m$ the Dirac mass.  Thus, seesaw mechanism could also be used to explains the smallness of the neutrino mass at the electro-weak energy scale.  The mass matrix model of a massive Majorana neutrino $M_{N}$ which is constrained by the solar and atmospheric neutrinos deficit and incorporate the seesaw mechanism and Peccei-Quinn symmetry have already been reported by Fukuyama and Nishiura \cite{Fukuyama}.  Using an $SU(2)_{L}\otimes SU(2)_{R}\otimes U(1)_{Y}$ gauge group with assumption that the form of the mass matrix is invariant under a cyclic permutation among fermions, and by choosing a specific neutrino mixing matrix, Koide \cite{Koide} obtained a unified mass matrix model for leptons and quarks that can be used to explain maximal mixing between $\nu_{\mu}$ and $\nu_{\tau}$ as suggested by the atmospheric neutrino data.

In this paper, we construct and evaluate the predictive power of neutrino mass matrices arising from seesaw mechanism with a heavy Majorana neutrino mass matrix $M_{N}$ subject to texture zero and invariant in form under a cyclic permutation to the solar neutrino mixing phenomena. This paper is organized as follows: In Section 2, we determine the possible patterns for heavy Majorana neutrino mass matrices $M_{N}$ subject to texture zero and investigate it whether invariant under a cyclic permutation.  The $M_{N}$ matrices which are invariant in form under a cyclic permutation will be used to obtain a neutrino mass matrices $M_{\nu}$ via seesaw mechanism. In Section 3, we evaluate and discuss the predictive power of the resulting neutrino mass matrices $M_{\nu}$ against experimental results.  Finally, in Section 4 we give the conclusion.

\section{Texture Zero and Invariant Under a Cyclic Permutation}\label{secII}

According to the seesaw mechanism, the neutrino mass matrix $M_{\nu}$ is given by
\begin{eqnarray}
M_{\nu}\approx -M_{D}M_{N}^{-1}M_{D}^{T},
 \label{eq:6}
\end{eqnarray}
where $M_{D}$ and $M_{N}$ are the Dirac and Majorana mass matrices respectively.  If we take $M_{D}$ to be diagonal, then the pattern of the neutrino mass matrix $M_{\nu}$ depends only on the pattern of the $M_{N}$ matrix.  From eq. (\ref{eq:6}), one can see that the pattern of the $M_{N}^{-1}$ matrix will be preserved in $M_{\nu}$ matrix when $M_{D}$ matrix is diagonal.

Following standard convention, let us denote the neutrino current eigenstates coupled to the charged leptons by the $W$ boson as $\nu_{\alpha} (\alpha=e,\mu,\tau)$, and the neutrino mass eigenstates as $\nu_{i} (i=1,2,3)$, then the mixing matrix $V$ takes the form presented in eq. (\ref{eq:5}).  As we have stated explicitly above, we use the seesaw mechanism as the responsible mechanism for generating neutrino mass.  The Majorana mass term in Lagrangian is given by
\begin{eqnarray}
L=-\nu_{\alpha}M_{\alpha\beta}C\nu_{\beta}+h.c.
 \label{eq:6a}
\end{eqnarray}
where $C$ is the charge conjugation matrix.  For the sake of simplicity, we will assume that $CP$ is conserve such that $M_{\alpha\beta}$ is real.  Within this simplification, the neutrino mass matrix $M_{\alpha \beta}$ is diagonalized by the matrix $V$ as
\begin{eqnarray}
V^{T}M_{\alpha\beta}V=\bordermatrix{& & &\cr
&m_{1} &0 &0\cr
&0 &m_{2} &0\cr
&0 &0 &m_{3}\cr}.
 \label{eq:6b}
\end{eqnarray}

From recent experimental results, the explicit values of the mixing matrix moduli $V$ are given by \cite{Gonzales-Garcia}
\begin{eqnarray}
\left|V\right|=\bordermatrix{& & &\cr
&0.79-0.88 &00.47-0.61 &<0.20\cr
&0.19-0.52 &0.42-0.73 &0.58-0.82\cr
&0.20-0.53 &0.44-0.74 &0.56-0.81\cr}.
 \label{eq:6c}
\end{eqnarray}
According to the requirement that the mixing matrix $V$ must be orthogonal, for the first approximation, we can take the mixing matrix $V$ to be \cite{Harrison}
\begin{eqnarray}
V=\bordermatrix{& & &\cr
&\sqrt{2/3} &\sqrt{1/3} &0\cr
&-\sqrt{1/6} &\sqrt{1/3} &-1/\sqrt{2}\cr
&-\sqrt{1/6} &\sqrt{1/3} &1/\sqrt{2}\cr}.
 \label{eq:6d}
\end{eqnarray}

We can obtain the neutrino mass matrix $M_{\nu}$ by using the relation
\begin{eqnarray}
M_{\nu}=V(V^{T}M_{\alpha\beta}V)V^{T}.
 \label{eq:6e}
\end{eqnarray}
If we substitute Eqs. (\ref{eq:6b}) and (\ref{eq:6d}) into Eq. (\ref{eq:6e}), then we obtain
\begin{eqnarray}
M_{\nu}=\bordermatrix{& & &\cr
&(2m_{1}+m_{2})/3 &(m_{2}-m_{1})/3 &(m_{2}-m_{1})/3\cr
&(m_{2}-m_{1})/3 &(m_{1}+2m_{2}+3m_{3})/6 &(m_{1}+2m_{2}-3m_{3})/6\cr
&(m_{2}-m_{1})/3 &(m_{1}+2m_{2}-3m_{3})/6 &(m_{1}+2m_{2}+3m_{3})/6\cr}.
 \label{eq:6f}
\end{eqnarray}
Inspecting neutrino mass matrix $M_{\nu}$ in Eq. (\ref{eq:6f}), one can rewrite it in general pattern as follow
\begin{eqnarray}
M_{\nu}=\bordermatrix{& & &\cr
&P &Q &Q\cr
&Q &R &S\cr
&Q &S &R\cr}.
 \label{eq:6g}
\end{eqnarray}

If the general neutrino mass matrix $M_{\nu}$ in Eq.(\ref{eq:6g}) is taken as the neutrino mass matrix arising from seesaw mechanism, then we can write $M_{\nu}$ as
\begin{eqnarray}
M_{\nu}=\bordermatrix{& & &\cr
&P &Q &Q\cr
&Q &R &S\cr
&Q &S &R\cr}=M_{D}M_{N}^{-1}M_{D}.
 \label{eq:6h}
\end{eqnarray}
From Eq.(\ref{eq:6h}), and putting $M_{D}$ matrix to be diagonal
\begin{eqnarray}
M_{D}=\bordermatrix{& & &\cr
&m_{1}^{D} &0 &0\cr
&0 &m_{2}^{D} &0\cr
&0 &0 &m_{3}^{D}\cr},
 \label{eq:6i}
\end{eqnarray}
with the constraint that $m_{2}^{D}=m_{3}^{D}=m^{D}$ at high energy ($CP$ is conserved), then the pattern of $M_{N}^{-1}$ matrix is preserved in $M_{\nu}$ \cite{Ma}.  Thus, the $M_{N}^{-1}$ matrix patterns can be written as
\begin{eqnarray}
M_{N}^{-1}=\bordermatrix{& & &\cr
&A &B &B\cr
&B &C &D\cr
&B &D &C\cr},
 \label{eq:6j}
\end{eqnarray}
and then the $M_{N}$ reads
\begin{eqnarray}
M_{N}=\frac{1}{A(C^2-D^2)+2B^2(D-C)}\bordermatrix{& & &\cr
&C^2-D^2 &B(D-C) &B(D-C)\cr
&B(D-C) &AC-B^2 &-AD+B^2\cr
&B(D-C) &-AD+B^2 &AC-B^2\cr}.
 \label{eq:6k}
\end{eqnarray}

Now, we are in position to impose the requirement of texture zero into $M_{N}$ matrix.  The $M_{N}$ matrix in Eq. (\ref{eq:6k}) will have texture zero if one or more of the following constraints are satisfied: (i) $C=-D$, (ii) $AD-B^2=0$, (iii) $AC-B^2=0$, and (iv) $B=0$.  If $M_{N}$ matrix in Eq. (\ref{eq:6k}) has one or more of its entries to be zero (texture zero), then $M_{N}^{-1}$ matrix has one or more of its $2\times 2$ sub-matrices with zero determinants.  The texture zero of the mass matrix indicates the existence of additional symmetries beyond the Standard Model of Particle Physics of electro-weak interaction.  After lenghty calculations, we obtained eight possible patterns of heavy Majorana $M_{N}$ matrix with texture zero as one can read in \cite{Damanik}.  The eight possible patterns of $M_{N}$ heavy Majorana neutrino matrices read
\begin{eqnarray}
M_{N}=\bordermatrix{& & &\cr
&0 &a &a\cr
&a &b &c\cr
&a &c &b\cr},M_{N}=\bordermatrix{& & &\cr
&a &b &b\cr
&b &c &0\cr
&b &0 &c\cr},M_{N}=\bordermatrix{& & &\cr
&a &b &b\cr
&b &0 &c\cr
&b &c &0\cr},\nonumber \\M_{N}=\bordermatrix{& & &\cr
&a &0 &0\cr
&0 &b &c\cr
&0 &c &b\cr},M_{N}=\bordermatrix{& & &\cr
&0 &a &a\cr
&a &b &0\cr
&a &0 &b\cr},M_{N}=\bordermatrix{& & &\cr
&0 &a &a\cr
&a &0 &b\cr
&a &b &0\cr},\nonumber\\M_{N}=\bordermatrix{& & &\cr
&a &0 &0\cr
&0 &b &0\cr
&0 &0 &b\cr},M_{N}=\bordermatrix{& & &\cr
&a &0 &0\cr
&0 &0 &b\cr
&0 &b &0\cr}.\ \ \ \ \ \ \ \ \ \ \ \ \ \ \ \ \ \ \ \ \ \ \ \ \ \ \ \ \ 
 \label{eq:16}
\end{eqnarray}

By checking the invariant form of the resulting neutrino mass matrices $M_{N}$ with texture zero under a cyclic permutation, we found that there is no $M_{N}$ matrix with texture zero invariant under a cyclic permutation.  But, with additional assumption, there is a possibility to put the $M_{N}$ matrices with texture zero to be invariant under a cyclic permutation, those are the $M_{N}$ matrices with the patterns
\begin{eqnarray}
M_{N}=\bordermatrix{& & &\cr
&a &0 &0\cr
&0 &b &0\cr
&0 &0 &b\cr},
 \label{eq:inv1}
\end{eqnarray}
and
\begin{eqnarray}
M_{N}=\bordermatrix{& & &\cr
&0 &a &a\cr
&a &0 &b\cr
&a &b &0\cr},
 \label{eq:inv2}
\end{eqnarray}
if we impose an additional assumption that $a=b$.  With the above assumption, we finally obtained two of the $M_{N}$ matrices to be invariant under a cyclic permutation. The two $M_{N}$ matrices which invariant under a cyclic permutation are
\begin{eqnarray}
M_{N}=a\bordermatrix{& & &\cr
&1 &0 &0\cr
&0 &1 &0\cr
&0 &0 &1\cr},
 \label{eq:inv3}
\end{eqnarray}
and
\begin{eqnarray}
M_{N}=a\bordermatrix{& & &\cr
&0 &1 &1\cr
&1 &0 &1\cr
&1 &1 &0\cr}.
 \label{eq:inv4}
\end{eqnarray}

By substituting Eqs. (\ref{eq:6i}), (\ref{eq:inv3}) and (\ref{eq:inv4}) into Eq. (\ref{eq:6}), then we obtain the neutrino mass matrices ($M_{\nu}$) to be
\begin{eqnarray}
M_{\nu}=-\frac{1}{m_{N}}\bordermatrix{& & &\cr
&(m_{1}^{D})^2 &0 &0\cr
&0 &(m^{D})^2 &0\cr
&0 &0 &(m^{D})^2\cr},
 \label{eq:inv5}
\end{eqnarray}
with mass eigenvalues
\begin{eqnarray}
m_{1}=-\frac{(m_{1}^{D})^2}{m_{N}},\ \  m_{2}=-\frac{(m^{D})^2}{m_{N}},\ \  m_{3}=-\frac{(m^{D})^2}{m_{N}},
 \label{eq:eigenvalue1}
\end{eqnarray}
and
\begin{eqnarray}
M_{\nu}=\frac {1}{m_{N}^{'}}\bordermatrix{& & &\cr
&(m_{1}^{D})^{2} &-m_{1}^{D}m^{D} &-m_{1}^{D}m^{D}\cr
&-m_{1}^{D}m^{D} &(m^{D})^{2} &-(m^{D})^2\cr
&-m_{1}^{D}m^{D} &-(m^{D})^2 &(m^{D})^2\cr},
 \label{eq:inv6}
\end{eqnarray}
with mass eigenvalues
\begin{eqnarray}
m_{1}=\frac{(m_{1}^{D})^2-m_{1}^{D}\sqrt{(m_{1}^{D})^2+8(m^{D})^2}}{2m_{N}^{'}},\nonumber\\ m_{2}=\frac{(m_{1}^{D})^2+m_{1}^{D}\sqrt{(m_{1}^{D})^2+8(m^{D})^2}}{2m_{N}^{'}},\nonumber\\ m_{3}=\frac{2(m^{D})^2}{m_{N}^{'}},\ \ \ \ \ \ \ \ \ \ \ \ \ \ \ \ \ \ \ \ \ \ \ \ \ \ \ \ \ \ \ \ 
 \label{eq:eigenvalue2}
\end{eqnarray}
where $m_{N}=a$ and $m_{N}^{'}=2a$ respectively. The neutrino mass matrix in eq.(\ref{eq:inv5}) will give $\Delta m_{32}^2=0$, then it fails to predict the neutrino mixing phenomena on the atmospheric neutrino sector at electroweak energy scale.  Thus, we obtain only one neutrino mass matrix $M_{\nu}$ (Eq. (\ref{eq:inv6})) arising from seesaw mechanism with heavy Majorana mass matrix subject to texture zero and invariant under a cyclic permutation that can be used to account the neutrino mixing for both solar and atmospheric neutrinos.

The form of neutrino mass matrix in Eq. (\ref{eq:inv6}) has also obtained by Ma \cite{Ma03} from the general neutrino mass matrix in the flavor basis (where the charged-lepton mass matrix is diagonal) by using simple interchange discrete symmetry
\begin{eqnarray}
\nu_{e}\rightharpoonup \nu_{e}, \ \ \nu_{\mu}\rightleftharpoons \nu_{\tau}.
\end{eqnarray}
%
\section{Discussions}\label{secIII}

Without imposing an additional assumption to the $M_{N}^{-1}$ matrices, we have no $M_{N}$ matrix invariant under a cyclic permutation. But, by imposing an additional assumption that $M_{N}^{-1}$ matrices have at least one of its $2\times2$ sub-matrices having zero determinant, then we have two $M_{N}$ matrices which invariant under a cyclic permutation. 

It also apparent that heavy Majorana mass matrix $M_{N}$ subject to texture zero in seesaw mechanism will give naturally a neutrino mass matrix without additional requirement that $M_{\nu}$ is invariant under a cyclic permutation.  This fact can be read in $M_{\nu}$ matrices arise from seesaw mechanism for the cases that $M_{N}$ matrix having one, two, or three of its entries to be zero in Eq. (\ref{eq:16}).   The $M_{N}$ matrices which having three zero entries will lead to the tri-maximal mixing.  One of the $M_{N}$ matrices that having three zero entries (all of the diagonal $M_{N}$ are zero) and invariant under a cyclic permutation gives a neutrino mass matrix $M_{\nu}$ with neutrino masses in normal hierarchy that can be used to explain the experimental result qualitatively.  If heavy neutrino Majorana $M_{N}$ matrix has six of its entries to be zero (all of the $M_{N}$ off-diagonal entries to be zero), then we obtain $M_{\nu}$ matrix as shown in Eq. (\ref{eq:inv5}) arises from seesaw mechanism which can not be used to explain neutrino mixing phenomena. The $M_{N}$ matrix with pattern
\begin{eqnarray}
M_{N}=\bordermatrix{& & &\cr
&a &0 &0\cr
&0 &0 &b\cr
&0 &b &0\cr},
 \label{eq:bimax}
\end{eqnarray}
can produces $M_{\nu}$ that can be used to explain neutrino bi-maximal mixing.

From Eq. (\ref{eq:eigenvalue2}), we obtain the neutrino masses $\left|m_{1}\right|=\left|m_{2}\right|\approx \sqrt{2}m_{1}^{D}m^{D}/m_{N}^{'}$ and $m_{3}=2(m^{D})^2/m_{N}^{'}$ for the case $m_{1}^{D} << m^{D}$, $\left|m_{1}\right|\approx (m^{D})^2/m_{N}^{'}, m_{2}\approx 2(m^{D})^2/m_{N}^{'}$, and $m_{3}=2(m^{D})^2/m_{N}^{'}$ for the case $m_{1}^{D}\approx m^{D}$, and $m_{1}\approx 0, m_{2}\approx (m_{1}^{D})^2/m_{N}^{'}$, and $m_{3}=2(m^{D})^2/m_{N}^{'}$ for the case $m_{1}^{D}>> m^{D}$.  The case $m_{1}^{D} << m^{D}$ gives a normal hierarchy  for neutrino masses and can predict qualitatively the masses square differences in Eqs. (\ref{eq:1}) and (\ref{eq:3}).

\section{Conclusion}\label{secIV}
The eight possible patterns of heavy Majorana neutrino matrix $M_{N}$ subject to texture zero could produce the neutrino mass matrices $M_{\nu}$ using seesaw mechanism as one can read in Ref.\cite{Damanik} and it could account for bi- and tri-maximal mixing in neutrino sector without additional assumption that these matrices are invariant in form under a cyclic permutation.   By evaluating the $M_{N}$ invariance under a cyclic permutation, we found that there is no $M_{N}$ matrix to be invariant in form.  But, by imposing an additional assumption that the $M_{N}^{-1}$ matrix has at least one of its $2\times2$ sub-matrices having zero determinant, we obtain two of the $M_{N}$ matrices invariant under a cyclic permutation. One of the two $M_{N}$ matrices (which are invariant under a cyclic permutation) could produces the neutrino mass matrix $M_{\nu}$ (Eq. (\ref{eq:inv6})) from the seesaw mechanism that can be used to explain the neutrino mixing phenomena for both solar and atmospheric neutrinos.
\section*{Acknowledgments}
The first author would like to thank to Directorate for Higher Education Ministry of National Education (Dikti Depdiknas) for a BPPS Scholarship Program.

\end{document}